%Paper: hep-th/9207076
%From: preskill@theory3.caltech.edu (John Preskill)
%Date: Wed, 22 Jul 92 16:14:40 PDT

\input phyzzx
% Titlepage macros
\hoffset=0.2truein
\voffset=0.1truein
\hsize=6truein
\def\TITLEPAGE{\frontpagetrue}
\def\CALT#1{\hbox to\hsize{\tenpoint \baselineskip=12pt
        \hfil\vtop{
        \hbox{\strut hep-th/9207076}
        \hbox{\strut CALT-68-#1}}}}

\def\CALTECH{
        \address{California Institute of Technology,
Pasadena, CA 91125}}

\def\AUTHOR#1{\vskip .2in \centerline{#1}}

\def\ABSTRACT#1{\vskip .2in \vfil \centerline{\twelvepoint
\bf Abstract}
        #1 \vfil}
\def\ENDTITLEPAGE{\vfil\eject\pageno=1}
\def\m{{\bf \mu}}

\def\n{{\bf n}}
\def\d{{\bf d}}
\def\k{{\bf k}}
\def\r{{\bf r}}
\def\z{{\bf z}}
\def\Z{{\bf Z}}
\def\y{{\bf y}}
\def\x{{\bf x}}
\TITLEPAGE
\CALT{1805}
\bigskip           %if titlepage is 2 lines use \break
\titlestyle {Simple Quantum Systems in Spacetimes with Closed Timelike
Curves\foot{Work supported in part by the U.S. Dept. of Energy
under Contract no. DEAC-03-81ER40050.}}
\AUTHOR{H. David Politzer}
\CALTECH
\ABSTRACT{Three simple examples illustrate properties of path integral
amplitudes in fixed background spacetimes with closed timelike curves:
non-relativistic potential scattering in the Born approximation is
non-unitary, but both an example with hard spheres and the exact
solution of a totally discrete model are unitary.}

\ENDTITLEPAGE

\eject

Path integral or sum-over-histories quantum mechanics has been proposed
as a possible means of generating a self-consistent dynamics in the
presence of closed timelike curves (CTC's).$^{1}$  It has been
argued,$^{2-5}$ however, that for interacting systems such evolution
(e.g., from before a compact region with CTC's to after) is necessarily
non-unitary.  (Such non-unitarity would present dire, though perhaps not
insuperable,$^{2,3,6}$ obstacles to the interpretation of the
predictions of such mechanics.)   The purpose of this paper is to offer
three very simple examples to illuminate the issue of unitarity.

The first example, considered in Section II, is non-relativistic
particles that scatter via a real potential in  the Born
approximation as one particular particle traverses a simply specified time
machine that
defines the compact CTC region.  This is just a variant of Boulware's
calculation$^{4}$ of relativistic particles in a Gott spacetime, but it
has two virtues: ~1)  One can carry the calculation to the end and close
a logical loophole left open in Ref. 4.  2)  The example is so simple
that there is no mystery or subtlety as to how the non-unitarity
arises.  This calculation is also an example of the general analysis of
perturbation theory given in Ref. 3 and agrees with those arguments.

Inspired by Thorne and Klinkhammer,$^{5}$ I consider in Section III WKB
hard sphere quantum mechanics with the same simple time machine as
defined in Section II.  If one includes ~a)  all numbers of
self-encounters and ~b) an excluded volume effect on the
``disconnected'' graphs, the amplitudes are Galilean covariant
(otherwise they would not be) and unitary.  In fact, they are equal to
the non-interacting amplitudes for particles traversing the time
machine.  Hence, there is unitarity but no net interaction with the
potentially dangerous time travelers.

In Section IV, I turn to a minimal discrete model that can be solved by
enumeration of configurations.  It is intended to be a cartoon of a
general, non-linear quantum field theory with a specified compact
region of CTC's.  The local field variable is reduced to two possible
values; the spatial positions inside the time machine are reduced to one
location; all spatial positions outside are likewise one location; and
time is discrete.  The model is presumably no more a free field theory
than the general, non-critical Ising model.  A nearest neighbor action
which gives unitary time evolution on the normal, flat spacetime lattice
generates a different but unitary evolution from before the CTC to
after.

If the non-unitarity of perturbation theory$^{3,4}$ is, indeed, generic,
then it behooves us to understand what is special about the
non-perturbative systems discussed in Sections III and IV.  Regrettably,
no such explanation is offered at present.

The philosophy and motivation of current investigations of time machines
is to seek out whether some fundamental principle forbids their
existence or whether they are physically realizable.  Even failing that
thus far, these questions offer a challenging context to test and stretch
our understanding of gravity and quantum mechanics.  For now, we begin with
a little background.

\noindent{\bf I.  Background}

It is not known at present whether a compact region containing CTC's can
arise in the context of classical gravitation.$^{7}$  Microscopic
versions may exist as quantum fluctuations of spacetime.  Alternatively,
large CTC regions may exist as relics of the quantum gravity epoch of
the Big Bang.

Entertaining the existence of time machines as worthy of consideration,
one is faced with two paradoxes of classical particle mechanics:  a
collision may render an ``earlier'' portion of a trajectory as
inconsistent with the collision itself, and given initial conditions may
correspond to several trajectories that satisfy the classical equations
of motion.$^1$  An action formulation allows one to consider only those
trajectories that are globally self-consistent.  And a quantum action
principle gives an interpretation to the multiple classically allowed
trajectories.  Each is a stationary point of the action, but all paths
are added coherently with the appropriate phase.  (An inherently quantum
mechanical singularity in the stress tensor does apparently develop just
before the first CTC's.$^{8}$  This is thought by some to signal a
backreaction which, handled consistently, may forbid the formation of
CTC's.  However, the strength of the singularity is sufficiently weak
that the relevant distance scales are so small as to require a quantum
gravity analysis, and the sign of the effect, opposite for fermions and
bosons, is not understood.)

Without a globally defineable time sequence or foliation, there is no
Hamiltonian evolution in the presence of CTC's and, hence, no obvious
reason for unitarity of evolution from before to after the CTC region.
Nevertheless, free particle systems yield unitary evolution.$^{9}$
Viewed in terms of particle trajectories, this unitarity relies on
cooperation between the different numbers of windings through the time
machine.  (See Appendix B for a sketch of a proof in this language.)

It has been argued that interacting systems are generally not unitary
(even if the same local action on a foliable spacetime yields unitary
Hamiltonian evolution).  This paper considers three examples.  The
background spacetimes are chosen by {\it fiat}, there being no known
``realistic'' models with compactly generated CTC's.  For simplicity,
the spacetimes are locally flat, with all the curvature located at
singular points.

Non-unitary amplitudes may still be used to generate relative
probabilities for sequences  of events or observations.$^{3,6,}$
However, there is a consequent acausality in that construction because
the geometry of all future CTC's have an effect, in principle, on
current observations.

\noindent{\bf II.  The Born Approximation}

The background spacetime is defined as follows.  Figure 1 illustrates
the construction in $1 + 1$ dimensions.   In the flat space, whose
points are labeled $(z,t)$, the heavy lines centered at $z = y_0$ and of
length $Y$ at $t = 0$ and $t = T$ are identified so that along them the
region immediately before $t = 0$ connects smoothly to that after $t =
T$, while the region immediately before $t = T$ connects smoothly to
that after $t = 0$.  In the new spacetime, we preserve the original
local direction of time and can use the old coordinates to label
points.  The handle thus formed contains the CTC's.

The spacetime is flat except for the two singular points at the handle
ends that have excess angles of $2\pi$.  The topological theorems
regarding compactly generated CTC regions$^{10}$ are satisfied, albeit
somewhat singularly.  For example, the Cauchy or chronology horizon
(i.e., the onset of the CTC region) is shrunk to the singular points.
(If one were to smooth out the curvature over a finite region, the
Cauchy horizon would be the first lightlike curves that circle the
handle.)  Also, there exist the isolated geodesics that enter but do not
exit the CTC region (or {\it vice versa}); these are the limiting case
of trajectories that enter (or exit)  the CTC region at velocity $v$ as $v
\rightarrow 0$.

As the Born approximation always has zero radius of convergence in $1 +
1$ dimensions (because all attractive potentials have at least one bound
state), I will use $3 + 1$ dimensions for an explicit example.  A
generalization of the spacetime defined above is clear: a compact region
of space (e.g., sphere) centered on ${\bf y}_0$ and of volume $Y^3$ is
identified at $t = 0$ and $t = T$ in flat spacetime analogously to the
$1 + 1$ dimensional case to define the time machine.

The dynamics is that of non-relativistic bosons ($\hbar = m = 1$).  Their
flat space, free propagator is
$$	K_f (\z_2, t_2; \z_1, t_1) = \cases {[2\pi i (t_2 -
t_1)]^{-3/2} \exp (i(\z_2 - \z_1)^2/(2(t_2 - t_1)) & $t_2 > t_1$\cr
\delta^3 (\z_2 - \z_1) & $t_2 = t_1$\cr
0 & $t_2 < t_1\,\, .$\cr}\eqno (2.1)$$
The particles interact via a real two-particle potential $\lambda V(r)$,
chosen (for simplicity) to depend on the magnitude of the two-particle
separation $r$.

Let $K(\z_2, t_2; \z_1, t_1)$ be the amplitude corresponding to all
paths that begin at $(\z_1, t_1)$ and end at $(\z_2, t_2)$ including all
windings of the machine, self-scatterings, and scatterings off closed
loops within the machine.  And define the coefficients of the $\lambda$
expansion of $K$ by $K = K_0 + \lambda K_1 + \lambda^2 K_2 + ... ~$.
There are disconnected paths (winding around the machine) that
contribute a common factor to $K$, to the before-to-after
vacuum-to-vacuum amplitude, and to all other amplitudes.  Hence, they
are to be divided out and, in practice for the present context, ignored.

{\it Important notation convention}:  I adopt the following convention
to indicate the allowed domains for spatial position coordinates:
Points restricted to lie {\it within} the identified volume $Y^3$ will
be labeled $\y$ (with subscripts and primes).  Points restricted to be
{\it outside} will be labeled $\x$.  Unrestricted positions are $\z$.
Position integrals are to be taken over the thus implicitly defined
ranges.

Unitarity of single particle evolution from before to after the time
machine would require
$$	\int d\x K(\x, T; \x_1, 0) K^* (\x, T; \x'_1, 0) = \delta^3
(\x_1 - \x'_1) \,\, . \eqno (2.2)$$
The free particle amplitude $K_o$ (which includes all windings of the
time machine) is unitary in this sense.  So next consider the ${\cal
O}(\lambda)$ contribution to eq. (2.2). Is
$$	\int d\x [K_1 (\x, T; \x_1,0) K_0^* (\x, T; \x'_1, 0) + K_0 (\x,
T; \x_1, 0) K_1^* (\x, T; \x'_1, 0)] = 0 \,\, ? \eqno (2.3)$$

To evaluate eq. (2.3) in closed form, I make the further simplification
that $T \gg Y^2$ and consider points $\x_1$ and $\x'_1$ such that $T \gg
(\x_1^{(')} - \y_0)^2$.  For such a large $T$ machine, the leading
contribution comes from the minimal number of windings (as discussed in
Appendix B).

The minimal, i.e., one, winding contribution to $\int d\x K_1 (\x, T;
\x_1,0) K_o^* (\x, T; \x'_1, 0)$, defined to be $A(\x_1, \x'_1)$, is
illustrated in Fig. 2.  Lines with arrows pointing up (down) are factors
of $K_f^{(*)}$, the dashed line is a factor of $V(|\z_2 - \z_1|)$, the
heavy horizontal lines define the time machine, and the factors must be
integrated over $0 < t < T$ and all $\z_1, \z_2, \y$, and $\x$.

Boulware$^{4}$ notes that for a field theoretic local $\lambda
\varphi^4$ interaction, the ${\cal O}(\lambda)$ contribution is of the
form of a particle scattering off an effective potential given by
$\lambda K(\z, t; \z, t)$ (which, naively, is the density of particles
looping the machine).  However, this is clearly a complex, oscillatory
function of $\z$ and $t$.  Since unitarity of potential scattering
requires a real potential, Boulware concludes that  unitarity is
violated.

In the present case, the analysis can be carried a bit further to
address the following two issues:  Since the question of unitarity
cannot be posed without integrating over all $\z_1, \z_2$ and $t$ between
$0$ and $T$, is it possible that the net integrated effect is, in fact,
unitary?  And if not, is the non-unitarity trivial, e.g., is the
amplitude unitary up to an overall factor, which could then be
reabsorbed into the measure?  (This latter possibility is realized in
the example of Section III.)  The answers here are no as found in
Appendix A.  For simplicity, use the time machine as the origin of
coordinates, i.e., take $\y_0 = 0$.  Then the $T \gg \x_1^{(')2} \gg Y^2$
amplitude is
$$	A(\x_1, \x'_1) \simeq {Y^3\over8\pi} (2\pi i T)^{-3/2}
e^{i(\x'_{1} - \x_{1})^{2}/2T} \left[{\x'_1 \cdot (\x_1 - \x'_1)\over |\x_1
- \x'_1|^3} ~W \left({\x'_1 \cdot (\x_1 - \x'_1)\over |\x_1 -
\x'_1|}\right) + \right.$$
$$	\left.+ {\x'_1 \cdot (\x_1 - \x'_1)\over x'^3_1} ~W
\left({\x'_1 \cdot (\x_1 - \x'_1)\over x'_1}\right) \right]
\,\, . \eqno (2.4)$$
where $W(r)$ is a complex linear functional of $V(r)$ defined in eq.
(A.4), which satisfies $Re W(r) = V(r)$ and $W(-r) = W(r)^*$.  Quite
generally, then, $A(\x_1, \x'_1) + A^* (\x'_1, \x_1) \not= 0$.

\noindent{\bf III.  Hard Spheres}

Quantum billiards or impenetrable spheres can be treated in a WKB
approximation because their interaction, rather than being smooth on the
scale of a wavelength, can be treated as a boundary condition.$^{5}$
(The ``approximation'' is thus exact, in the sense it is exact for free
particles.)  I consider here a single particle's traversal of the same
time machine as described in Section II.  In particular, I consider
paths of initially compact wavepackets whose spread is small compared to
the hard sphere diameter.  The packets traverse the time machine in a
proper time sufficiently small that wavepacket spreading can be ignored;
this can be guaranteed for all numbers of windings and self-collisions
by suitable choice of initial conditions and ratio of the hard sphere
size to the size of the time machine.

The amplitudes are given by a factor of $i \exp \{i S_{classical}\}$,
where $S_{classical} = {1\over 2} (\Delta \z/\Delta t)^2$, for each
collisionless leg of the journey and a $(-i)$ for each collision.

It is instructive to go to a moving frame rather than the machine rest
frame.  Let the center of the identified region at $t = T$ be $\y'_0$
and that at $t = 0$ be $\y_0$ such that $\y'_0 - \y_0 = \d$.  Paths with
$0,1,$ and $2$ windings are shown in Fig. 3.

The action corresponding to a single winding for a particular collision
point is ${1\over 2} (\d/T)^2$.  For one winding, the integral over the
possible locations of the collision gives a factor of $L/D$ where $L$ is
the length of the projection of the no collision path onto the
identified volumes.  (In $1 + 1$ dimensions, $L$ is simply $Y$.)  $D$ is
the billiard ball diameter.  For $n$ windings, the factor is
$(L/D)^n/n$!.  Hence, the sum of amplitudes for such paths that go from
$(\x, 0)$ to $(\x', T)$ with windings $n = 0, 1,2 ...$ is (taking $L \gg
D$)
$$	\sum_{n = 0}^\infty e^{{i\over 2} ({\x' - \x\over
T})^2} {i^{n + 1}\over n!} \left({L\over D}\right)^n \left(
e^{{i\over 2} ({\d\over T})^2}\right)^n$$
$$	= i e^{{i\over 2} ({\x' - \x\over T})^2} \exp\left\{i {L\over D}
e^{{i\over 2} ({\d\over T})^2}\right\} \,\, . \eqno (3.1)$$
This, itself, is not unitary for $\d \not= 0$.  Nor, however, is it
Galilean covariant.  What is missing is a correct account of the
disconnected paths.

When the initial collisionless path from $\x$ to $\x'$ does not traverse
the positions in the time machine, the machine is threaded by closed
loops.  The amplitude for these loops is the coherent sum over all
numbers of loops, integrated over their allowed trajectories as
restricted by the excluded volume effect of the impenetrable spheres.
The sum of these closed loops amplitudes is also the before-to-after
vacuum-to-vacuum amplitude.

When the initial collisionless path does traverse the machine, each of
the paths included in eq. (3.1) excludes a volume for possible closed
loop, disconnected paths, i.e., the velocity $\d/T$ paths that pass
through the collisionless path.  However, the factor thus lost from the
completely disconnected volume is precisely the factor acquired by
summing the possible collisions as in eq. (3.1).  Hence, the product of
the amplitude in eq. (3.1) with the allowed disconnected loops is equal
to the product of the unitary collisionless amplitude with the hard
sphere vacuum-to-vacuum amplitude.  All of the apparent non-unitarity
and frame (or $\d$) dependence resides in a common factor of all
amplitudes, which is the naive before-to-after vacuum-to-vacuum
amplitude.  This factor, however, is unobservable and is properly
divided out everywhere.

\noindent {\bf IV.  An Ising Model}

Finally, I consider a totally discrete model that can be solved by
enumeration of the finite number of configurations.  The analog flat
spacetime, depicted in Fig. 4a, consists of a $2 \times m$ lattice,
i.e., with two spatial positions and $m$ times.  At each lattice site
$\n$, there is a two-valued field $s(\n) = \pm 1$.  The ``path
integral'' is
$$	Z = \sum_{s(\n)} e^{i \alpha \sum\limits_{\n,\m} (s(\n + \m) -
s(\n))^2} \,\, , \eqno (4.1)$$
where $\m$ runs over the two positive unit vectors.  The choice $\alpha
= \pi/8$ yields a unitary $4 \times 4$ transfer matrix that relates the
four possible $s$ configurations at a given time to those at the next
time.

The time machine is defined by reidentifying two of the timelike links
as indicated in Fig. 4b.  The amplitude of interest is the $4 \times 4$
matrix for the sum over all intermediate time configurations with a
particular initial configuration (immediately preceding the CTC) and a
particular final configuration (immediately following the CTC).  I have
done the sums for systems with $m = 3, 4$ and $5$.  In each of these
cases, the amplitude differs from the analogous flat spacetime system
but is, nevertheless, unitary.

\noindent {\bf V.  Conclusion}

The non-unitarity of interacting particle propagation across a compact
region of spacetime with closed timelike curves (CTC's) is demonstrated
with an exceedingly simple, non-relativistic example.  In a particular
limit of the parameters, all integrals can be evaluated
for arbitrary incoming states.  This was not done in previous analyses.
An analogous issue arrises in the general relativistic perturbation
theory analysis of Ref. 3.  There the non-unitarity is demonstrated by
identifying combinations of propagator functions that are non-zero in
the presence of CTC's but which would have integrate to zero against
general state functions were the theory unitary.  It is not immediately
obvious that the state functions form a complete set with respect to the
relevant integrals.

The non-unitarity of perturbation theory is presumably generic.
However, analogous calculations in two non-perturbative examples do not
exhibit non-unitarity at the same level.  This challenges us to better
understand the issues.  Free theories are unitary, presumably, because
particles that do go back in time still cannot influence anything in the
past, and they themselves eventually propagate into the future because
quantum diffusion prevents a particle from time cycling indefinitely
with a non-vanishing probability.  In the hard sphere example considered
here, there certainly are interactions, e.g., two incoming wavepackets
could scatter off each other, and there certainly is time travel, i.e.,
of the type performed by free particles, which definitely alters their
trajectories even after the CTC region.  But the particular time machine
considered here appears to have the property that the multiple classical
alternatives allowed by having both interactions and CTC's  sum to
something equivalent to  having no interactions.

The discrete model considered here is not a free field theory in that it
is not linearly coupled harmonic oscillators because of the restriction
$s = \pm 1$.  Its simplicity allows an exact solution but precludes much
in the way of interpretation.

A better understanding of each of these examples would be very welcome.

\noindent {\bf Acknowledgements}

Several valuable conversations with J. Preskill and S.A. Ridgeway are
gratefully acknowledged.

\noindent {\bf Appendix A}

The leading large $T$ behavior of the ${\cal O}(\lambda)$ unitarity
violation discussed in Section II can be evaluated in
three steps.  First consider the unitarity of the Born approximation in
flat spacetime for a single particle scattering off a potential.  Then
generalize to two-to-two particle scattering.  And, finally, modify the
latter to fit the time machine boundary conditions.

Let $B(\z_0, \z'_0)$ be the amplitude depicted in Fig. A.1, i.e.,
including integrals over $\z, \z'$ and $t$:
$$	B(\z_0, \z'_0) = - i \int d \z d \z' \int_0^T dt K_f (\z, t; \z_0,
0) V (\z) K_f (\z', T; \z, t) K^*_f (\z', T; \z'_0, 0) . \eqno (A.1)$$
Free particle unitarity reduces this to
$$	B(\z_0, \z'_0) = - i \int d\z \int_0^T dt K_f (\z, t; \z_0,0)
V(\z) K^*_f (\z, t; \z'_0, 0)$$
$$	= {-i\over (2\pi)^3} \int_0^T {dt\over t^3} \exp \{i (z_0^2 -
z'^2_0)/(2t)\} \tilde{V} \left({\z'_0 - \z_0\over t}\right) \,\, . \eqno
(A.2)$$
The second form uses the explicit form for $K_f$, and $\tilde{V}$ is the
Fourier transform of $V$.  Note that if $V$ is real, $\tilde{V} (\k) =
\tilde{V}^* (-\k)$, and then $B(\z_0, \z'_0) + B^* (\z'_0, \z) = 0$,
which is the statement of ${\cal O}(\lambda)$ unitarity of potential
scattering.

The $t$ integral in eq. (A.2) can be expanded about the large
$T$ limit (changing variables to $k = 1/t$)
$$	B (\z, \z') = {-i\over (2\pi)^3} \int_0^\infty k dk \exp \{i
(z^2 - z^2_0) k/2\} V ((\z' - \z)k) + {i\over 16\pi^3} ~ {\tilde{V}(0)\over
T^2} + ... \,\, . \eqno (A.3)$$
The $k$ integral is reminiscent of the radial part of a spherically
symmetric Fourier transform.  In particular, if we restrict $V(\r)$ to
real functions that depend only on $r$, the magnitude of $\r$, and
define the function $W(r)$ by
$$	W(r) \equiv - {i\over 2\pi^2 r} \int_0^\infty k~ dk~ e^{ikr}
\tilde{V}(k) \,\, , \eqno (A.4)$$
then $Re ~W(r) = V(r)$ (which is symmetric under $r \rightarrow - r$)
while $Im~W(r)$ is antisymmetric in $r$.  In terms of $W$,
$$B(\z, \z') = {1\over 8\pi} ~ {z^2 - z'^2\over |\z - \z'|^3} ~W
\left({z^2 - z'^2\over 2|\z - \z'|}\right) + {i \tilde{V} (0)\over
16\pi^3 T^2} + ... \,\, . \eqno (A.5)$$

For two-to-two scattering, the analogous amplitude $B(\z_1, \z_2, \z'_1,
\z'_2)$, depicted in Fig. A.2, is given by
$$	B(\z_1, \z_2, \z'_1 \z'_2) = \delta^3 (\Z - \Z') [B(\z, \z') +
B(\z, - \z')] \eqno (A.6)$$
where
$$	\eqalign{\Z &= (\z_1 + \z_2)/2\cr
	\z &= \z_1 - \z_2\cr} \eqno (A.7)$$
and the same definitions hold for the primed coordinates.  Eq. (A.6)
reflects that the problem is separable into free center of mass motion
and scattering in the relative coordinate.  Unitarity to ${\cal O}
(\lambda)$ is again clearly satisfied.

The amplitude we need to test unitarity to ${\cal O} (\lambda)$ in the
time machine is $A(\z_1, \z_2$, $\z'_1, \z_3)$, illustrated in Fig. A.3.
It is simply related to the $B$'s using the unitarity of $K_f$, which
implies that $K_f$ is the inverse of $K^*_f$.  Hence
$$	A(\z_1, \z_2, \z'_1, \z_3) = \int d \z'_2 \delta^3 (\Z - \Z')
[B(\z, \z') + B(\z, - \z')] K_f (\z_3, T; \z'_2, 0) \,\, . \eqno (A.8)$$
This final integral over $\z'_2$ is trivial because of the
$\delta$-function.

The one particle amplitude relevant to the time machine is (recalling
the implicit ranges of coordinates $\x, \y,$ or $\z$ defined in Section
II)
$$	\eqalign{A(\x_1, \x'_1) &= \int d\y A( \x_1, \y, \x'_1, \y)\cr
	&\simeq Y^3 A(\x_1, \y_0, \x'_1, \y_0) \,\, \cr} \eqno (A.9)$$
where the second form is the leading term for small $Y$.  $A(\x_1,
\x'_1) + A^* (\x'_1, \x_1) \not= 0$ (the actual expression is given in
eq. (2.4) using eq. (A.7) and the $\delta$-function) which implies there
is no unitarity.

\noindent {\bf Appendix B}

I address briefly free particle unitarity in the $T \gg Y^3$ time
machine and for general $T$ and discuss some aspects of windings in
general and small $Y$.

The expansion of the free particle amplitude $K_0 (\x', T; \x, 0)$ from
before to after the time machine in terms of numbers of windings and the
flat space free propagator $K_f$ looks like (remembering $\x$'s are
outside and $\y$'s are inside the $Y^3$ volume)
$$	K_0 (\x', T; \x, 0) = K_f (\x', T; \x, 0) + \int d\y K_f (\x', T;
\y, 0) K_f (\y, T; \x, 0)$$
$$	+ \int d\y d\y' K_f (\x', T; \y, 0) K_f (\y, T; \y', 0) K_f (\y',
T; \x, 0) + ... \,\, . \eqno (B.1)$$
To check unitarity, we replace the $\x''$ integral in $ \int d \x'' K_0
(\x'', T; \x, 0) K_0^* (\x'', T; \x',0)$ with an integral $d\z''$, i.e.,
as if $\x''$ ran over the full range, minus an integral $d\y''$.  The
integral $d\z''$ always yields a $\delta$-function because of $K_f$
unitarity.  To leading order for $T \gg Y^3$, the non-unitarity of $K_f$
when restricted to end points outside the time machine is canceled by a
contribution from the one-winding term of $K_0$.  (The evaluation is
straightforward.)  All effects of higher windings (and a residual
non-unitarity of the one-winding term) are down by $Y^3/T^{3/2}$.

For arbitrary $T$, the unitarity of $K_0$ with the same time machine can
be demonstrated using the same expansion.$^{11}$  Each successively higher
winding restores the unitarity of the one fewer winding amplitude but
introduces its own non-unitarity.  Hence, one must sum all windings to
recover unitarity.

A general amplitude written in terms of $K_f$'s, i.e., before
integrating over any spacetime coordinates, will have various $\y_i$
arguments.  The integrals $\int d \y_i f(\y_i)$ can be replaced by $Y^3
f (\y_0)$ in the small $Y$ limit as long as all the $\y$'s are
independent.  If, however, some $\int d\z$ yields a $\delta^3 (\y_i -
\y_j)$, then there is one fewer factor of $Y^3$ than given by counting
the $\y$'s.  This is essential to the free particle case discussed
above.  The only such $\z$ integral that occurs in the generalization of
the calculation of Appendix A to higher winding numbers comes from the
factor $\int d\x K_0^* (\x, T; \x'_1 0) K_0 (\x, T, \z, t)$.  For this
integral, the following identity holds:
$$	\int d \x K_0^* (\x, T; \x',0) K_0 (\x, T, \z, t) = K_f^* (\z,
t; \x', 0) \,\, , \eqno (B.2)$$
which follows from free particle unitarity.  The same identity with the
$K_0$'s replaced by $K_f$'s was used to get eq. (A.2).  Hence, adding
all possible windings to the paths for this portion of the calculation
has no net effect.  Finally, adding higher windings to the other
segments of the paths of the calculation of Appendix A indeed gives
extra factors of $Y^3$.  Hence, the leading small $Y$ behavior is given
by the minimal winding amplitude.

\noindent {\bf Figure Captions}

\item{1.}  A time machine in $1 + 1$ dimensions.  The heavy lines are
identified such that the shaded regions join smoothly to each other and
are disconnected from the likewise joined crosshatched regions.

\item{2.}  The ${\cal O}(\lambda)$ propagator unitarity testing
amplitude, $A(\x_1, \x'_1)$.  The lines pointing upward denote factors
of $K_f$, downward $K_f^*$, and dashed horizontal $V$; the solid
horizontal lines denote the time machine as in Fig. 1.

\item{3.}  Hard sphere WKB trajectories through the CTC region with $0,
1$ and $2$ self-collisions.

\item{4.}  a)  A flat $2 \times 5$ lattice spacetime, with arrows
indicating the positive sense of the timelike links.

\item{~}   b)  A $2 \times 5$ spacetime with a closed timelike curve.
\bigskip

\item{A.1} The ${\cal O} (\lambda)$ unitarity amplitude for potential
scattering, $B (\z_0, \z'_0)$.

\item{A.2}  The ${\cal O} (\lambda)$ unitarity amplitude for two
particle potential scattering, $B (\z_1, \z_2$, $\z'_1, \z'_2)$.

\item{A.3}  The truncated two particle amplitude, $A (\z_1, \z_2, \z'_1,
\z_3)$.

\noindent {\bf References}

\item{1.}  F. Echeverria, G. Klinkhammer, and K.S. Thorne, Phys. Rev. D
{\bf 44}, 1077 (1991).

\item{2.}  D. Deutsch, Phys. Rev. D {\bf 44}, 3197 (1991).

\item{3.}  J.L. Friedman, N.J. Papastamatiou, and J.Z. Simon, U. of
Wisconsin-Milwaukee preprint, ``Failure of unitarity for interacting
fields on spacetimes with closed timelike curves'' (1992).

\item{4.}  D.G. Boulware, U. of Washington preprint UW/PT-92-04 (1992).

\item{5.}  G. Klinkhammer and K.S. Thorne, unpublished.

\item{6.}  J.B. Hartle, U.C. Santa Barbara preprint UCSBTH-92-04 (1992).

\item{7.}  e.g. M.S. Morris, K.S. Thorne, and U. Yurtsever, Phys. Rev.
Lett. {\bf 61}, 1446 (1988); U. Yurtsever, Class. Quant. Grav. Lett.
{\bf 7}, L251 (1990); G. Klinkhammer, Phys. Rev. D {\bf 43}, 2542
(1991); V.P. Frolov, Phys. Rev. D {\bf 43}, 3878 (1991); R.M. Wald and
U. Yurtsever, Phys. Rev. D {\bf 44}, 403 (1991).

\item{8.}  S.-W. Kim and K.S. Thorne, Phys. Rev. D {\bf 43}, 3929
(1991).

\item{9.}  e.g. J.L. Friedman, N.J. Papastamatiou, and J.Z. Simon, U.
of Wisconsin-Milwaukee preprint WISC-MIL-91-TH-17 (1991).

\item{10.}  S. Hawking, Phys. Rev. D (to be published).

\item{11.}  J. Preskill, private communication.
\bye